\documentclass[conference]{IEEEtran}
\IEEEoverridecommandlockouts
\usepackage{cite}
\usepackage{amsmath,amssymb,amsfonts}
\usepackage{algorithmic}

\usepackage{graphicx}
\usepackage{textcomp}
\usepackage{xcolor}
\def\BibTeX{{\rm B\kern-.05em{\sc i\kern-.025em b}\kern-.08em
    T\kern-.1667em\lower.7ex\hbox{E}\kern-.125emX}}
    
\usepackage{url}  

\title{Privacy-Preserving Identifier Checking in 5G}

\author{\IEEEauthorblockN{Marcel D.S.K. Gr{\"a}fenstein}
\IEEEauthorblockA{Technische Universit{\"a}t Dresden\\
 Dresden, Germany\\
E-Mail: grafensteinmarcel20@gmail.com}
\and
\IEEEauthorblockN{Stefan K{\"o}psell}
\IEEEauthorblockA{Technische Universit{\"a}t Dresden / Barkhausen Institut\\
Dresden, Germany\\
E-Mail: stefan.koepsell@tu-dresden.de\\
E-Mail: stefan.koepsell@barkhauseninstitut.org}\\

\and
\IEEEauthorblockN{Maryam Zarezadeh}
\IEEEauthorblockA{Barkhausen Institut\\
Dresden, Germany\\
E-Mail: maryam.zarezadeh@barkhauseninstitut.org}}

\usepackage[most]{tcolorbox}
\usepackage{tikz}
\usetikzlibrary{positioning, shapes, arrows.meta}
\usepackage{tabularx}
\usepackage{booktabs}
\usepackage{pifont} 
\usepackage{caption}
\usepackage{float}
\usepackage{acronym}
\usepackage{multirow} 
\usetikzlibrary{arrows.meta,positioning,fit,shapes.misc,calc}

\usepackage[linesnumbered,ruled,vlined]{algorithm2e}
\usepackage{todonotes}
\setuptodonotes{inline}

\begin{document}

\maketitle

\begin{abstract}
Device identifiers like the International Mobile Equipment Identity (IMEI) are crucial for ensuring device integrity and meeting regulations in 4G and 5G networks. However, sharing these identifiers with Mobile Network Operators (MNOs) brings significant privacy risks by enabling long-term tracking and linking of user activities across sessions. In this work, we propose a privacy-preserving identifier checking method in 5G. This paper introduces a protocol for verifying device identifiers without exposing them to the network while maintaining the same functions as the 3GPP-defined Equipment Identity Register (EIR) process.
The proposed solution modifies the PEPSI protocol [USENIX, 2024] for a Private Set Membership (PSM) setting using the BFV homomorphic encryption scheme. This lets User Equipment (UE) prove that its identifier is not on an operator’s blacklist or greylist while ensuring that the MNO only learns the outcome of the verification. The protocol allows controlled deanonymization through an authorized Law Enforcement (LE) hook, striking a balance between privacy and accountability.
Implementation results show that the system can perform online verification within five seconds and requires about 15 to 16 MB of communication per session. This confirms its practical use under post-quantum security standards. The findings highlight the promise of homomorphic encryption for managing identifiers while preserving privacy in 5G, laying the groundwork for scalable and compliant verification systems in future 6G networks.
\end{abstract}

\section{Introduction}
\label{sec:introduction}
Mobile networks rely on persistent and globally unique device identifiers—Permanent Equipment Identifiers (PEIs)—to ensure integrity, access control, and regulatory compliance. In 3GPP systems, the primary PEI is the 15-digit IMEI, composed of a Type Allocation Code (TAC), a device-specific Serial Number (SNR), and a Check Digit (CD)~\cite{3gpp_5g_2024-4,3gpp_numbering_2024,3gpp_system_2024}. IMEIs enable UE recognition across LTE, 5G, and NG-RAN and support cross-domain device management.
5G and LTE verify PEIs through the Equipment Identity Register (EIR), classifying devices as \textit{WHITELISTED}, \textit{BLACKLISTED}, or \textit{GREYLISTED}~\cite{3gpp_procedures_2024,3gpp_5g_2024-1}. Whitelisted devices receive normal service; blacklisted ones (typically stolen) are denied access; greylisted ones remain under investigation. Serving Networks also obtain the PEI during roaming~\cite{cox_architecture_2021}, enabling theft prevention, fraud mitigation, and support for legal investigations~\cite{gsma_ts06_2024}.
Despite this operational role, permanent identifiers pose a significant privacy risk. Disclosing a stable, globally unique PEI at every access attempt allows long-term cross-session correlation—even across pseudonymous identities such as the SUPI. In modern 5G environments with extensive analytics, precise location data, and multi-access edge processing, these risks scale further and contradict 5G’s \emph{Zero Trust} principles. As networks evolve toward data-driven 6G architectures, GDPR-aligned operation requires reducing identifier exposure while preserving lawful accountability.

\emph{Problem Statement.}
Current EIR-based verification mechanisms rely on plaintext transmission of the PEI:
whenever a device attaches or reattaches to the network, its identifier is directly disclosed to the MNO or SN~\cite{3gpp_5g_2024-1,3gpp_procedures_2024}.
Even if transported securely, the MNO learns the identifier in clear form and may associate a persistent device identity with session metadata, mobility patterns, and subscriber interactions.
This creates a technically simple basis for user profiling, long-term tracking, and cross-network linkability.
The core challenge is therefore to enable \emph{identifier verification without identifier disclosure}.
A privacy-preserving mechanism must allow the MNO to determine whether a device belongs to a specific operator-maintained set--such as the Blacklist or Greylist--while learning nothing about the identifier unless a legally authorized process requires disclosure.
Achieving this under strict 5G constraints on latency, scalability, and regulatory compliance defines the central problem addressed in this work.

\emph{Proposed Approach Overview.}
To mitigate these privacy risks, we propose a privacy-preserving PEI verification protocol based on homomorphically encrypted Private Set Membership (PSM). Our design adapts PEPSI~\cite{mahdavi_pepsi_2024} and uses the Brakerski--Fan--Vercauteren (BFV) scheme (Microsoft SEAL~4.1.2) to perform efficient encrypted equality checks over IMEIs under RLWE-based post-quantum security.
During verification, the UE encrypts its identifier and the MNO evaluates membership against the Blacklist and Greylist in parallel. The protocol outputs \emph{Not-Listed}, \emph{Listed}, or \emph{Non-Evaluable}, and reveals no identifier information to the MNO. A bounded randomization mechanism provides statistical privacy without risking ciphertext overflow, enabling large-scale evaluations.
For lawful accountability, the protocol includes a controlled \emph{Law-Enforcement (LE) hook}. The UE provides an additional $\mathrm{Enc}_{\mathsf{LE}}(\mathrm{PEI})$ ciphertext; if a Greylist match is found, the MNO forwards this ciphertext and session metadata to LE, allowing authorized deanonymization without exposing the identifier to the operator.
The approach remains compatible with standard 3GPP registration and authentication flows, including emergency-registration scenarios. Its primary change is replacing plaintext EIR queries--the main source of identifier leakage--with homomorphically protected membership checks.

\emph{Contributions.}
This work provides the following scientific and technical contributions: a) A 5G-aligned privacy-preserving identifier verification protocol derived from a PEPSI-inspired PSM construction and instantiated with BFV homomorphic encryption. b) A selective, auditable Law-Enforcement hook enabling deanonymization of greylisted devices under judicial authorization while preserving privacy by default. c) A complete implementation and experimental evaluation using Microsoft SEAL~4.1.2 with realistic 5G-scale identifier sets. d) A quantitative analysis of runtime, communication, and noise-budget trade-offs, demonstrating the post-quantum practicality of homomorphic identifier verification. e)  A bounded randomization mechanism for reliable homomorphic masking without overflow, ensuring correctness and scalability during large set evaluations. f) An empirical evaluation of privacy–efficiency trade-offs, highlighting how BFV parameter tuning affects decryptability, security margins, and suitability for future 6G frameworks.

\section{Related Work}
\label{sec:related-work}
The PEI is checked against operator-maintained EIRs in LTE and 5G~\cite{3gpp_system_2024,3gpp_procedures_2024,3GPP_TS29511_2023}. While this enables blocking or monitoring of listed devices, it exposes permanent identifiers to the operator and returns only a Blacklist/Greylist decision~\cite{3gpp_5g_2024-1}, leaving privacy unaddressed.
Cryptographic approaches have been proposed to hide identifiers during list checking. ZKP-based schemes (e.g., zk-SNARKs, zk-STARKs, Bulletproofs) can prove membership without revealing the queried value, but large, frequently updated lists make proving costly and require expensive setup or update procedures~\cite{el-hajj_evaluating_2024}, limiting their suitability for dynamic mobile systems. Private Set Membership (PSM) protocols such as PEPSI~\cite{mahdavi_pepsi_2024} offer a more practical alternative: they support dynamic updates, distributed list management, and low-latency checks at far lower overhead, and allow reuse of server-side material across updates.
Building on PEPSI~\cite{mahdavi_pepsi_2024}, we adapt its PSI mechanism into a PSM protocol aligned with the 5G system architecture, enabling an EIR-equivalent verification service without revealing the PEI. Our evaluation emulates large operator registries and shows that PSM naturally supports dynamic updates with near-constant client cost, bridging cryptographic feasibility and architectural requirements.

\emph{Telecom Identifiers and Lawful Accountability.}
Most privacy work in mobile networks targets subscriber identifiers in 5G-AKA~\cite{wang2021privacy,koutsos20195g,you20235g}, yet measurements still reveal linkability risks from paging, analytics, and virtualized cores~\cite{eleftherakis2024demystifying,scalise2025survey,saeed2022comprehensive}. Equipment identifiers (PEI/IMEI) in the EIR~\cite{3GPP_TS29511_2023} remain largely exposed. Existing proposals—PGPP~\cite{schmitt2021pretty}, P3LI5~\cite{intoci2023p3li5}, and regulatory systems such as CEIR~\cite{DoT_CEIR_2023v02} and the GSMA Device Registry~\cite{gsma_ts06_2024}—highlight the tension between traceability and privacy but rely on heavy cryptography or centralized LEA databases. Our work contributes a 5G-aligned, privacy-preserving PEI-verification mechanism based on PEPSI-PSM and BFV that preserves EIR functionality without exposing plaintext identifiers, while still enabling auditable, regulation-compliant LEA access.

\section{Preliminaries}
\label{sec:preliminaries}
This section introduces the key entities, sets, and cryptographic foundations that form
the basis of the proposed protocol.

\emph{System Entities and Sets.}
\label{sec:system_entities}
The protocol involves three entities: the UE, the MNO, and an LE agency.  
The UE holds a single permanent equipment identifier, denoted
$\mathcal{S}_{UE}=\{\mathrm{PEI}_{UE}\}$, which it must prove to be absent from the operator’s
lists through a PSM-based non-membership test.
The MNO maintains two disjoint sets of device identifiers:
a Greylist $\mathcal{G}$ containing PEIs subject to lawful monitoring,
and a Blacklist $\mathcal{B}$ containing PEIs blocked from network access.  
For verification, the MNO operates on the union
$\mathcal{S}_{MNO}=\mathcal{G}\cup\mathcal{B}$, although only the Blacklist is directly managed by
the operator, while the Greylist is maintained by the LE.
The LE does not participate in every protocol run but maintains a legally authorized subset of
greylisted identifiers, $\mathcal{S}_{LE}\subseteq\mathcal{G}$, and may perform deanonymization of
matching identifiers when required by lawful procedures.

\emph{Threat Model.}
\label{sec:threat_model}
We assume that each UE uses its correct PEI during protocol execution. Attacks such as IMEI cloning, spoofed identifiers, or semi-permanent PEIs target lower system layers and fall outside the privacy-preserving verification problem considered here.
We adopt a semi-honest adversary model: parties follow the protocol but may attempt to infer hidden information from their inputs, outputs, or ciphertexts. Because large-scale systems such as 5G cannot realistically exclude governmental oversight, we assume the LE to be trustworthy for identifier verification and operating within its jurisdiction. Techniques that hide PEIs even from LE agencies are orthogonal to this work. Privacy is preserved if the MNO learns only the membership result of the PEI query, while the LE performs deanonymization solely through authorized procedures.
We also consider limited \emph{evasive client behaviour} after the UE decrypts the masked PSM response. A UE that suspects it is listed may attempt to manipulate messages to hide its status; such behaviour constitutes a protocol deviation and is detectable during the MNO’s demasking and validation steps (see Sec.~\ref{sec:pm_protocol_arch_and_phases}). Full protection against PEI forgery or IMEI cloning remains outside our scope.

\emph{Homomorphic Encoding and Data Model.}
\label{sec:pre_he_dm}
The PSM operations use the BFV homomorphic encryption scheme (SEAL~4.1.2), providing 128-bit
post-quantum security under RLWE. BFV enables SIMD-style packing, allowing multiple comparison
slots to be encoded in a single ciphertext and evaluated component-wise.
In our adaptation, the PEPSI-based PSM computation outputs one encrypted bit per slot, indicating
whether the UE’s PEI matches an entry in the operator’s lists. During masking
(Section~\ref{sec:masking}), the MNO blinds all slots using a global multiplicative randomizer
$r_{1}$ and independent additive masks $r_{2,i}$, drawing fresh values for each PSM test.
Identifiers are first processed through permutation-based hashing (PBH), which reduces the IMEI
bit length $\lambda$ to an effective length $\bar{\lambda}$ and assigns them to slots. The slot
values are then encoded using constant-weight codewords (CWC) of Hamming weight $h$ and bit
length $l$, where $l$ is the smallest value satisfying
$
\binom{l}{h} \ge \bar{\lambda}.
$
These parameters together define the BFV layer ($N$, $q$, $t$), the PEPSI encoding layer
($\lambda$, $\bar{\lambda}$, $h$, $l$), and the masking layer ($r_{1}$, $r_{2,i}$,
$\mathrm{Result}_{i}$), eliminating the need for a separate parameter table.

\emph{Masking and Demasking.}
\label{sec:masking}
To prevent the UE from learning its membership status,
the MNO applies a two-layer masking scheme to the encrypted PSM result.
After the homomorphic set-membership test, each ciphertext
encodes per-slot results $\mathrm{Result}_i \in \{0,1\}$, representing match outcomes for each slot.
The MNO blinds these results using a global multiplicative randomizer and per-slot additive noise: $\mathrm{Masked}_i = r_1 \cdot \mathrm{Result}_i + r_{2,i} \pmod{t},$ where $r_1$ is reused across all slots (vector $[r_1,\ldots,r_1]$), while each $r_{2,i}$ is
sampled independently.
This ensures that the UE, after decryption, observes only random-looking values and cannot
infer whether or where a match occurred.
Any manipulation of ciphertexts or protocol messages alters the expected numeric range
or exceeds the permissible noise threshold, allowing the MNO to detect inconsistencies during demasking.


\section{Proposed Method}
\label{sec:proposed-method}
Building on the cryptographic foundations and threat model in Section~\ref{sec:preliminaries}, we present an adapted PEPSI-based PSM protocol for privacy-preserving identifier verification in 5G networks. The scheme applies PEPSI’s unbalanced PSI design to a single-identifier setting aligned with the 3GPP EIR procedure, enabling the MNO to check membership in its Black- and Greylists without learning the PEI while preserving EIR-equivalent functionality.

\emph{Design Objectives and Requirements.}
The protocol aims to let the MNO learn only whether a PEI is in $\mathcal{B}$ or $\mathcal{G}$, while the identifier itself remains hidden; only LE may recover it under judicial authorization. The design follows four goals: (i) \emph{privacy}, ensuring the MNO learns only the membership result; (ii) \emph{correctness and robustness}, detecting any ciphertext modification through demasking; (iii) \emph{lawful accountability}, via a controlled LE hook for greylist deanonymization (Sec.~\ref{sec:introduction}); and (iv) \emph{architectural alignment}, preserving the EIR logic in 3GPP~TS~29.511 without plaintext identifier exposure.

Quantitative and cryptographic constraints further shape the design. The protocol must support $\approx 2^{20}$ identifiers per list, evaluate $\mathcal{B}$ and $\mathcal{G}$ in a single encrypted query, and keep end-to-end verification below $5$\,s. We instantiate BFV at 128-bit RLWE security and use SIMD packing for constant per-entry complexity and efficient list updates. The UE performs only encryption, decryption, and light aggregation, while the MNO handles all homomorphic evaluation and masking.

\begin{figure}
    \centering
    \includegraphics[width=0.80\linewidth]{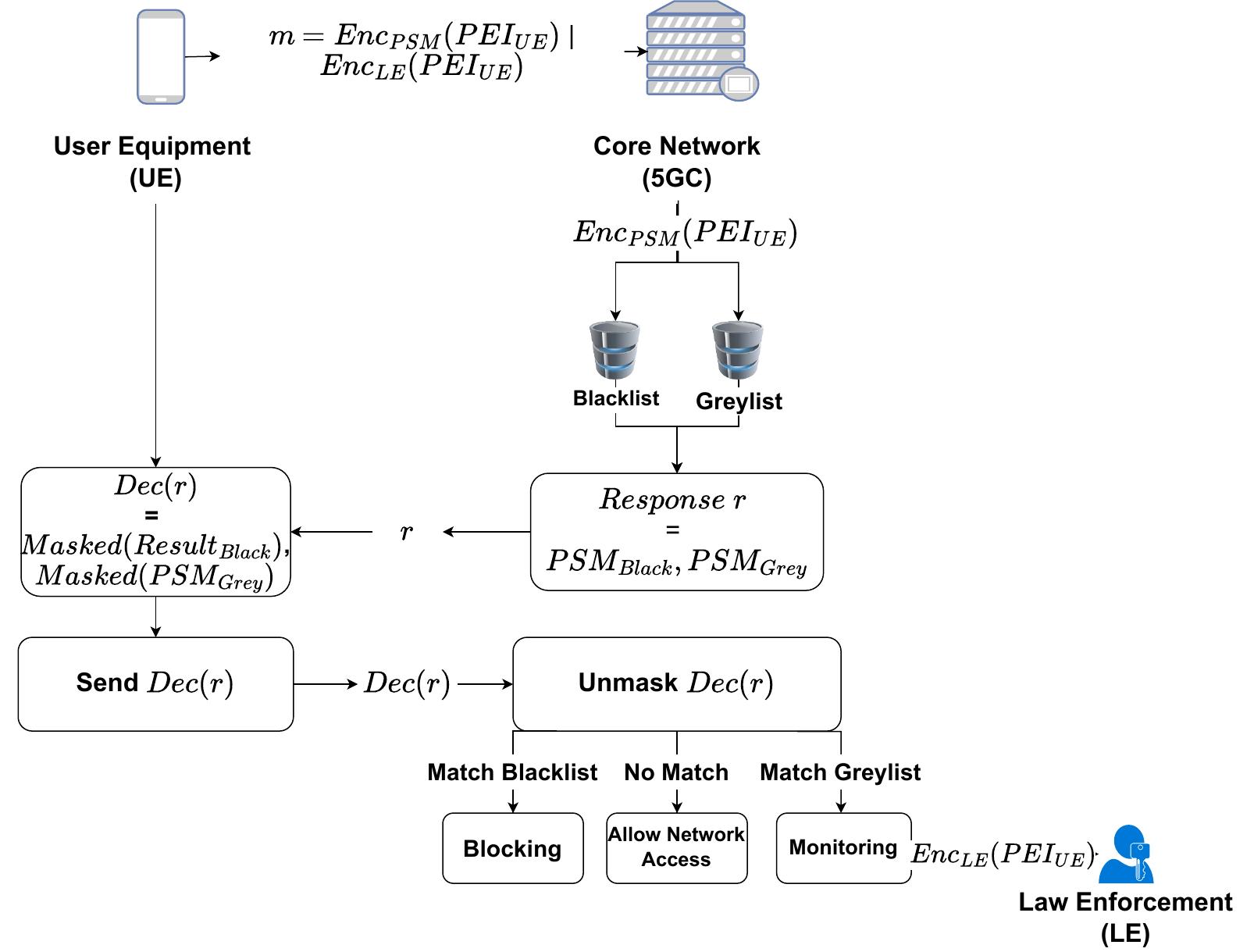}
    \caption{\small Identifier-verification protocol overview.}
    \label{fig:architecture}
\end{figure}
\subsection{Protocol Architecture and Phases}
\label{sec:pm_protocol_arch_and_phases}
The protocol involves three entities (Section~\ref{sec:system_entities})--UE, MNO, and LE--and proceeds in four phases: (1) setup and key distribution, (2) offline list preprocessing, (3) online verification and result return, and (4) audit and deanonymization. Figure~\ref{fig:architecture} summarizes the online phases (3) and (4).
In the \emph{Setup Phase.}, each UE instantiates the BFV context and generates the corresponding key material. The UE retains the secret decryption key and shares the public evaluation parameters with the MNO. These parameters are sufficient for the MNO to perform arithmetic operations on ciphertexts received from the UE without learning any plaintext information. In the \emph{Offline Phase.}, the MNO maintains its Blacklist $\mathcal{B}$ and Greylist $\mathcal{G}$, which is accessed under LE authorization. Both lists are stored in plaintext at the MNO and encoded as constant-weight codewords (CWCs) mapped to BFV plaintext polynomials, as introduced in Section~\ref{sec:pre_he_dm}. Since the lists remain local to the MNO, only entries that change must be re-encoded. As in PEPSI, each identifier is assigned to a SIMD slot, enabling slotwise equality tests between the UE's encrypted query and the list elements. The LE maintains a long-term asymmetric key pair $(pk_{LE}, sk_{LE})$ for authorized deanonymization requests; key management policies are decoupled from the identifier-verification protocol.

\begin{algorithm}
  \caption{\small Server computation and result extraction for PSISum -- adapted from Algorithm~5 in \cite{mahdavi_pepsi_2024}}
  \label{algo:psi_sum}
  \small
  \scalebox{0.9}{
  \begin{minipage}{\linewidth}
  \begin{algorithmic}[1]
    \STATE \textbf{Procedure} COMPUTE-PSI-SUM($ct_{c}, pt_{s}$)
    \STATE Homomorphically encrypted client identifier: $ct_{c}$
    \STATE Server list, e.g., Blacklist: $pt_{s}$
    \STATE Calculate the PSI table ($ct_{eq}[i][i']$) via slotwise comparison between all client and server elements
    \STATE \textbf{Additive random masking values: } $r_{2} \gets [r_{2,1}, r_{2,2}, \ldots, r_{2,N}]$
    \STATE \textbf{Global random multiplicative masking value: } $val_{s} \gets [r_{1}, r_{1}, \ldots, r_{1}]$ with $|val_{s}| = N$
    \STATE Adapted $ct_{sum} \gets \sum_{i'} val_{s}[i'] \cdot \sum_{i} ct_{eq}[i][i']$
    \STATE Ciphertext blinding: $ct_{res} \gets ct_{sum} + r$
    \STATE \textbf{return} $ct_{res}$
    \STATE \textbf{End Procedure}
  \end{algorithmic}
  \end{minipage}
  }
\end{algorithm}
\emph{Online verification and result return.}\label{sec:ovp}
The online phase is a 1.5-round interaction between UE and MNO.
\emph{UE request.}
When initiating network access, the UE encrypts its PEI under two public keys: its own BFV public key $pk$ for privacy-preserving verification and the LE public key $pk_{LE}$ for lawful access. Both ciphertexts are sent to the MNO.
\emph{MNO-side homomorphic evaluation.}
Upon receiving the request, the MNO matches the encrypted PEI against its plaintext lists $\mathcal{B}$ and $\mathcal{G}$. Following PEPSI, it constructs ciphertext--plaintext equality indicators over all list entries and applies an adapted PSISum computation that aggregates equality results in each slot and masks the outcome using slotwise additive noise and a global multiplicative randomizer, as summarized in Algorithm~\ref{algo:psi_sum}. The MNO sends the blinded ciphertext $ct_{res}$ back to the UE.

\emph{UE decryption and aggregation.}
After decryption, the UE obtains a vector of masked slot values $(y_0',\ldots,y_{N-1}')$. Returning the full vector would allow the MNO to demask each slot individually and learn the slot index of any match, partially revealing $PEI_{UE}$, because the MNO can assign every slot a subset of PEIs. To avoid this positional leakage and minimize communication, the UE instead computes an aggregated scalar
\begin{equation}
\label{eq:ptsummod}
pt_{sum} = \sum_{i=0}^{N-1} y_i' \bmod t
\end{equation}
and returns only this 64-bit value to the MNO.

\emph{MNO demasking and decision.}
Upon receiving $pt_{sum}$, the MNO subtracts the aggregated additive mask and compares the result against the global multiplicative randomizer $r_1$ and 0. This demasking step recovers a ternary outcome (Match, No-Match, or Protocol Deviation) without exposing the underlying PEI or list structure. Algorithm~\ref{algo:extract_info} formalizes the extraction logic. Depending on the result, the MNO authorizes network access, initiates monitoring for greylisted identifiers, or rejects the connection in the presence of a blacklist match or protocol deviation.

\begin{algorithm}
  \caption{\small Private Set Membership Test -- Extract Information (Server)}
  \label{algo:extract_info}
  \small
  \scalebox{0.9}{
  \begin{minipage}{\linewidth}
  \begin{algorithmic}[1]
    \STATE \textbf{Procedure} Serverside-PSM($pt_{\text{sum}}$)
    \STATE Receive modular slot sum $pt_{\text{sum}}$ from UE
    \STATE Additive server noise: $r_{2} \gets [r_{2,1}, r_{2,2}, \ldots, r_{2,n}]$
    \STATE Compute $R_{2} \gets \sum_{i=1}^{n} r_{2}[i]$
    \STATE Global multiplicative randomizer: $r_{1}$
    \IF{$pt_{\text{sum}} - R_{2} \equiv r_{1} \pmod t$}
      \STATE \textbf{return} Match
    \ELSIF{$pt_{\text{sum}} - R_{2} \equiv 0 \pmod t$}
      \STATE \textbf{return} No-Match
    \ELSE
      \STATE \textbf{return} Protocol Deviation
    \ENDIF
    \STATE \textbf{End Procedure}
  \end{algorithmic}
  \end{minipage}
  }
\end{algorithm}

\emph{Security Against Client Manipulation.}
\label{sec:sec_client_manip}
Beyond PEI confidentiality, the protocol also mitigates a restricted form of client-side manipulation that may occur after the UE decrypts the masked PSM result 
(cf.~Section~\ref{sec:pm_protocol_arch_and_phases}).
Here, a malicious UE may try to alter the returned value so that the MNO accepts an incorrect membership status during demasking.
After the homomorphic set-membership test, each plaintext slot encodes $\text{Result}_i \in \{0,1\}$, which the MNO blinds as
$
  \text{Masked}_i = r_1 \cdot \text{Result}_i + r_{2,i} \bmod t
$,
where $r_1 \neq 0$ is a global multiplicative randomizer and each $r_{2,i}$ is an independent additive mask in $\mathbb{Z}_t$. The UE decrypts only the masked values $y_i'$, learning neither $r_1$ nor the $r_{2,i}$.
In the aggregated variant, the UE computes
$
  pt_{\text{sum}} = \sum_{i=0}^{N-1} y_i' \bmod t
$
and returns only this value. Let $R_2 = \sum_{i=0}^{N-1} r_{2,i} \bmod t$. Then:  
(i) in the \emph{No-Match} case, $pt_{\text{sum}} \equiv R_2 \pmod t$;  
(ii) in the \emph{Match} case, exactly one index $j$ has $\text{Result}_j = 1$, giving $pt_{\text{sum}} \equiv R_2 + r_1 \pmod t$.  
During demasking, the MNO recomputes $R_2$ and checks if $pt_{\text{sum}} - R_2$ is $0$, $r_1$, or neither.
To force a No-Match outcome in the Match case, a UE must send $pt_{\text{sum}}'$ such that $pt_{\text{sum}}' \equiv R_2 \pmod t$, even though the true value satisfies $pt_{\text{sum}} \equiv R_2 + r_1 \pmod t$. This requires guessing $r_1$ and setting $pt_{\text{sum}}' = pt_{\text{sum}} - r_1 \bmod t$. Since $r_1$ is uniformly sampled in each run, a single forgery attempt succeeds with probability $1/t$. The same reasoning applies to forging a Match result from a genuine No-Match state.
As fresh masks are used across sessions, attempts are independent. For $n$ protocol executions, the probability of at least one successful forgery is at most $n/t$ by a union bound. Section~\ref{sec:evaluation} evaluates this bound for our chosen parameters.

\emph{Audit and Deanonymization Phase}\label{sec:audit_and_deanony.}
If a greylisted identifier triggers a monitoring condition, the MNO forwards the pre-attached $Enc_{LE}(\mathrm{PEI})$ to the competent LE authority, which decrypts this value using its private key $sk_{LE}$ and recovers the PEI associated with the greylist match. The deanonymization channel is strictly one-way: the LE receives $Enc_{LE}(\mathrm{PEI})$ only for UEs that triggered a lawful event, and the MNO never gains the capability to decrypt PEIs.
Practical aspects such as the transport of $Enc_{LE}(\mathrm{PEI})$, session identifiers, and logging formats vary between jurisdictions and deployments and are orthogonal to the privacy-preserving identifier check; they are left as implementation choices. The LE key pair $(pk_{LE}, sk_{LE})$ is independent from the MNO and UE key material, ensuring a clean cryptographic separation between operational verification and lawful identification domains. Since the encrypted PEI is transmitted only once during the online phase and never modified thereafter, this mechanism imposes negligible computational overhead and does not affect the latency of the main protocol, while providing an auditable means of lawful access.
\subsection{Complexity (Unbalanced Setting)}
\label{sec:complexity_summary}
We now summarize communication and online computation in the unbalanced setting ($m \ll n$).
\emph{Communication.} The communication pattern mirrors PEPSI's unbalanced case~\cite{mahdavi_pepsi_2024}: a request from UE$\to$MNO and a response from MNO$\to$UE. Under hashing-to-bins, the PEPSI request grows linearly in the client set size $m$. In the single-identifier 5G scenario ($m{=}1$), the UE sends a constant number of BFV ciphertexts determined by encoding and HE parameters, and the MNO returns one BFV ciphertext. Our 1.5-round adaptation adds a single 64-bit value from UE to MNO (Eq.~\eqref{eq:ptsummod}), which does not change the linear-in-$m$ behaviour~\cite[Sec.~6.1]{mahdavi_pepsi_2024} (see Tab.~\ref{tab:complexity-pepsi-vs-adapted}).
\begin{table}[t]
\centering
\caption{\small Communication complexity and rounds in the unbalanced setting ($m \ll n$).}
\label{tab:complexity-pepsi-vs-adapted}
\setlength{\tabcolsep}{3pt}
\renewcommand{\arraystretch}{1.12}
\begin{tabularx}{\columnwidth}{@{} l X c @{}}
\toprule
\textbf{Protocol}
& \textbf{Communication (asympt.)}
& \textbf{Rounds} \\
\midrule
PEPSI~\cite[Sec.~5.3]{mahdavi_pepsi_2024}
& $O(m)$
& 1 \\
Adapted PEPSI 5G (this work)
& $O(m){+}O(1)$
& 1.5 \\
\bottomrule
\end{tabularx}
\end{table}
\emph{Online computation.} The MNO's online work is dominated by homomorphic equality checks over all server slots, as in PEPSI Algorithm~2, yielding a runtime that grows linearly in the server set size $n$~\cite[Thm.~2]{mahdavi_pepsi_2024}. Subsequent Circuit-PSI steps (including PSISum) are comparatively cheap and return one ciphertext to the UE~\cite{mahdavi_pepsi_2024}. In the adapted protocol, the UE performs an $O(N)$ aggregation over decrypted slots and the MNO executes an additional host-level modular accumulation; for fixed BFV parameters (constant $N$), this contributes only an additive term and leaves the dominant complexity linear in $n$.

\section{Evaluation}
\label{sec:evaluation}
This section evaluates the implemented PEPSI-based PSM test (Fig.~\ref{fig:architecture}) in terms of runtime and communication overhead. We focus on verification against a Blacklist; Greylist checks run in parallel with identical complexity, so overall performance is dominated by the larger dataset, here the Blacklist. The list contains $2^{20}$ randomly generated 14-digit IMEIs (without check digits) to reduce the bitlength $\lambda$ and the effective bitlength~$\bar{\lambda}$.
The evaluation answers four questions: (Q1) Is the protocol fast enough for 5G-scale verification? (Q2) Which BFV parameters preserve sufficient noise budget? (Q3) What is the per-login communication cost? (Q4) Does masking remain correct under the chosen arithmetic constraints?
All experiments use Microsoft SEAL~4.1.2. Each configuration was executed five times on distinct IMEIs, reporting average, standard deviation, and maximum runtime, together with the PSM result. Each identifier is tested once to emulate realistic single-login behaviour in 5G attachment, leveraging protocol randomness rather than repeated identical trials.
The implementation distinguishes \emph{offline} and \emph{online} phases for both UE and MNO: offline phases handle preprocessing and fresh masking generation, while online phases perform the PSM evaluation and demasking.

\emph{Implementation Settings.} We instantiated the adapted PEPSI-based PSM protocol using the parameter-selection workflow of Mahdavi et al.~\cite{mahdavi_pepsi_2024}. The main tunables are the Hamming weight $h$, the polynomial modulus degree $N$, and the coefficient modulus~$q$. Following PEPSI, we encode 47-bit IMEIs with PBH, giving an effective bitlength $\bar{\lambda} = \lambda - \log_2(N)$, which reduces payload while preserving entropy.
We used the authors' script \texttt{find-comp-optimal-hw.py} to explore $(N, \bar{\lambda}, h)$ combinations. Each candidate was tested via the C++ reference implementation (10 runs), recording runtime and communication. Configurations with reconstructed bitlength $\lambda = \bar{\lambda} + \log_2(N) \neq 47$ were discarded to avoid entropy loss or padding. Among the valid points, we selected the configuration with the lowest runtime under acceptable communication cost.
Two options were competitive: (i) $h=7$, $\bar{\lambda}=34$ (slightly faster but with larger messages), and (ii) $h=8$, $\bar{\lambda}=34$ (moderate runtime, about 14.5\,s, with smaller requests of about 14.1\,MB). We adopt (ii). According to Table~3 in~\cite{mahdavi_pepsi_2024}, this corresponds to $h=8$, $N = 2^{13}$, $q \approx 2^{204}$, and a SEAL plaintext modulus $t = 1{,}032{,}193$ from \texttt{PlainModulus::Batching($2^{13}$, 20)}. This configuration provides roughly 128-bit classical security and serves as the baseline for our evaluation.

\emph{Baseline Results.} We first evaluated the baseline parameter set. Server-side offline computation dominated with about 81~s for both Blacklist and Whitelist runs, which is acceptable because it can be precomputed and cached. The online phase completed in roughly 4~s ($<5$~s), staying within typical latency bounds for 5G attachment, while client-side work remained negligible ($<50$~ms). However, none of the returned ciphertexts decrypted correctly: the noise budget was fully depleted during server-side homomorphic computation, so the configuration was performant but not decryptable.

\emph{Noise-Budget Analysis.} In BFV, each homomorphic operation increases ciphertext noise, and once the accumulated noise exceeds the modulus-dependent bound, decryption fails. For the baseline configuration, all tested combinations of Hamming weight and effective bitlength resulted in zero remaining noise budget, and tuning PEPSI-level parameters such as $l$ or $\bar{\lambda}$ alone did not restore correctness.
To address this, we examined three mitigation strategies: (i) scaling the PEPSI encoding parameters, (ii) segmenting the IMEI into a plaintext prefix and encrypted suffix, and (iii) increasing the BFV coefficient modulus independently of PEPSI. The first two approaches still yielded insufficient noise budget: even shortened identifiers and relaxed PEPSI-conform parameters did not lead to reliable decryption. Moreover, the segmentation strategy was discarded because exposing a plaintext prefix to the MNO would weaken unlinkability guarantees.

\emph{Whitelist and Blacklist Validation Using SEAL Default Settings.} Switching to SEAL's default coefficient modulus (\texttt{CoeffModulus::BFVDefault($2^{13}$)}) yielded a parameter set with $\approx$128-bit classical security that preserved a positive noise budget throughout the PSM test and restored correct decryption. The larger ciphertexts slightly increased the UE request size from about 14.8~MB to 16.4~MB.
In the Whitelist experiment (150 non-matching identifiers), all runs succeeded, confirming the correctness of the protocol logic for all $\text{IMEIs} \notin \text{Blacklist}$. Initial Blacklist runs, however, still showed sporadic failures despite sufficient noise budget, caused by overflow in the masking step: sampling $r_1 \in [1, t)$ and $r_{2,i} \in [0, t)$ allowed $r_1 + r_{2,i} \ge t$ and wrap-around. Restricting the ranges to $r_1 \in [1, \tfrac{t}{2}-1]$ and $r_{2,i} \in [0, \tfrac{t}{2}-1]$ (with $r_1 > 0$ to avoid cancelling matches) removed these overflows, after which all 150 Blacklist tests decrypted correctly.
\begin{table*}[t]
\centering
\caption{\small Communication overflow and online/offline times for Blacklist and Whitelist tests (bytes; ms). Means and SDs are computed from per-run totals to capture sub-metric covariance.
}
\label{tab:comm-time-black-white}
\scalebox{0.85}{
\begin{tabularx}{\textwidth}{@{} l *{3}{X} *{2}{X} *{2}{X} @{} }
\toprule
\multirow{2}{*}{\textbf{Test}} & \multicolumn{3}{c}{\textbf{Communication Overflow [B]}} & \multicolumn{2}{c}{\textbf{Offline time [ms]}} & \multicolumn{2}{c}{\textbf{Online time [ms]}} \\
\cmidrule(lr){2-4} \cmidrule(lr){5-6} \cmidrule(lr){7-8}
 & \textbf{Client request $\pm$ SD} & \textbf{Server response $\pm$ SD} & \textbf{Client response $\pm$ SD} & \textbf{UE offline $\pm$ SD} & \textbf{MNO offline $\pm$ SD} & \textbf{UE online $\pm$ SD} & \textbf{MNO online $\pm$ SD} \\
\midrule
Blacklist (runs: 150) & 16{,}442{,}917.41~B $\pm$ 414.60~B & 103{,}088.85~B $\pm$ 82.91~B & 8~B $\pm$ 0~B & 52.57~ms $\pm$ 1.70~ms & 87{,}163.03~ms $\pm$ 1{,}760.54~ms & 0.01~ms $\pm$ 0.08~ms & 4{,}091.89~ms $\pm$ 88.16~ms \\
Whitelist (runs: 150) & 16{,}442{,}936.41~B $\pm$ 424.25~B & 103{,}078.87~B $\pm$ 64.98~B & 8~B $\pm$ 0~B & 52.67~ms $\pm$ 1.83~ms & 86{,}482.85~ms $\pm$ 1{,}474.22~ms & 0.00~ms $\pm$ 0.00~ms & 4{,}078.99~ms $\pm$ 63.34~ms \\
\bottomrule
\end{tabularx}
}
\end{table*}

\emph{Client-side manipulation resistance.}
Using the masking construction from Section~\ref{sec:pm_protocol_arch_and_phases} and Section~\ref{sec:sec_client_manip}, we quantify the forgery probability under our chosen parameters. All masking is performed over $\mathbb{Z}_t$ with $t = 1{,}032{,}193$. For each query, the MNO samples a fresh global multiplicative mask $r_1 \leftarrow [1,\lfloor t/2 \rfloor - 1]$ and additive masks $r_{2,i} \leftarrow [0,\lfloor t/2 \rfloor - 1]$ for all slots. The effective multiplicative space is
$
  t_{\mathrm{eff}} = \lfloor t/2 \rfloor - 1 = 516{,}095.
$
After decryption, the UE sees only masked slot values and returns the aggregate $pt_{\text{sum}}$ (Eq.~\eqref{eq:ptsummod}); the MNO then demasks using $R_2$ and $r_1$. A Blacklisted UE attempting to evade detection must output a forged $pt_{\text{sum}}'$ that passes demasking without knowing $r_1$. Even assuming the UE somehow knows all additive masks and the unmasked aggregate, it must still guess $r_1$ from a space of size $t_{\mathrm{eff}}$. Thus, the success probability of a single forgery is
$
  P_{\text{hit}} \le 1/t_{\mathrm{eff}}
  = 1/516{,}095
  \approx 1.94\times 10^{-6}
  \approx 2^{-19},
$
which also bounds attempts to fabricate a Match from a genuine No-Match.
Since masks are re-sampled independently across sessions, past failures give no advantage. For $n$ manipulation attempts, the union bound gives $P_{\text{succ}} \le n/t_{\mathrm{eff}}$. Even one attempt per day over five years ($n=1{,}825$) yields
$
  P_{\text{succ}} \approx 3.54\times10^{-3},
$
i.e., below $0.36\%$ in total, with operational safeguards (logging, rate limits) further reducing practical risk. The masking therefore offers strong integrity protection against evasive UEs.
The bounded randomization slightly affects the slot-value distribution: Whitelist tests remain uniform (via PBH), while Blacklist tests become approximately triangular due to summing two uniform variables. Although repeated queries could in principle leak mild statistical cues, this does not affect the cryptographic soundness of the masking scheme.

\begin{figure}[!ht]
    \centering
    \includegraphics[width=\columnwidth]{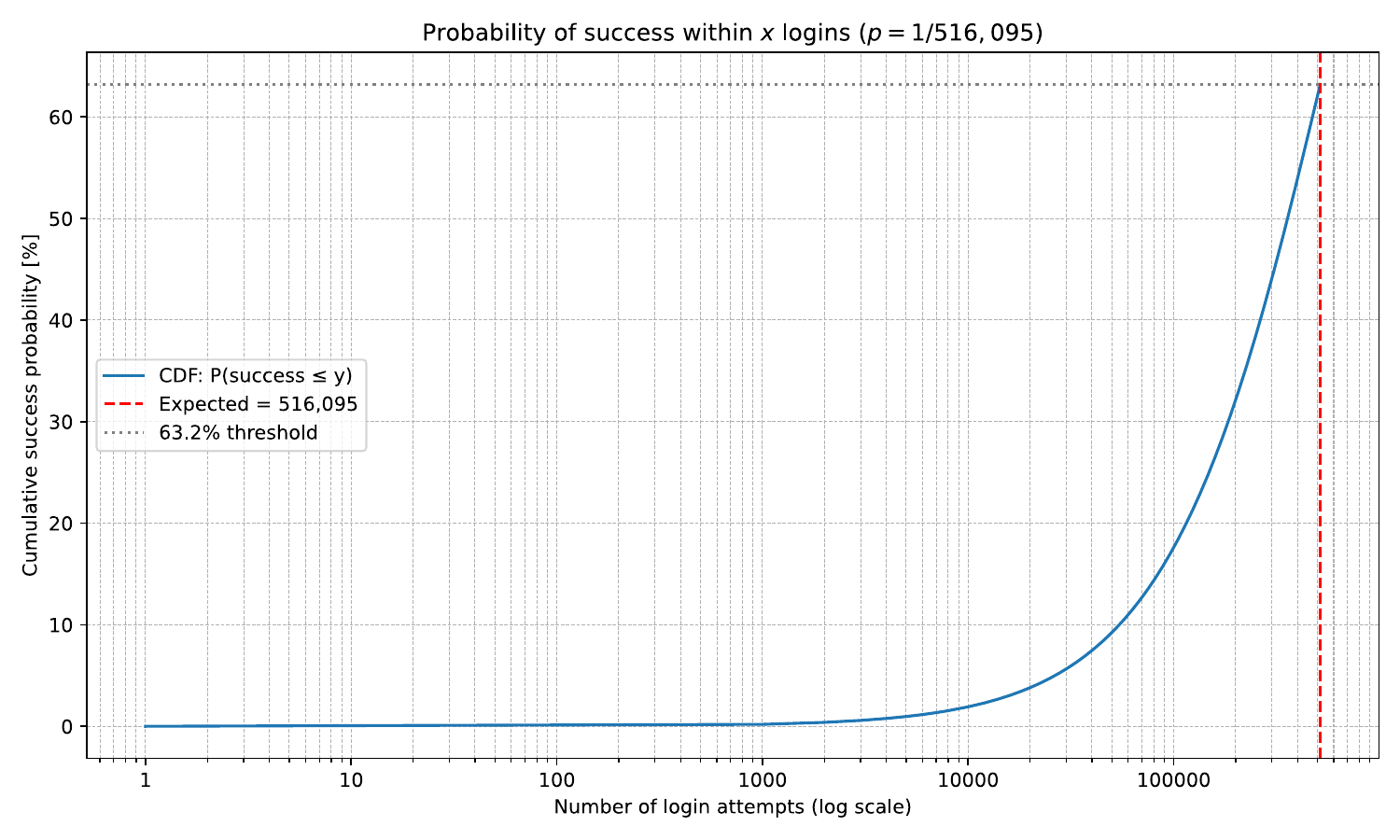}
    \caption{\small Cumulative success probability for a uniformly
    guessed value in a space of size $t_{\mathrm{eff}} = 516{,}095$
    over $x$ attempts (logarithmic scale). The expected number of
    attempts until the first success is $1/p = t_{\mathrm{eff}}$.}
    \label{fig:geom_dist_log}
\end{figure}
\emph{Discussion and Limitations.}
Our experiments indicate that PEPSI-based PSM provides practical privacy-preserving identifier verification under conservative SEAL parameters and bounded masking (Tab.~\ref{tab:comm-time-black-white}). The main limitations are:  
(i) \emph{communication} — each login requires about 16.4\,MB uplink and 103\,kB downlink;  
(ii) \emph{parameterization} — original PEPSI settings caused decryption issues, while SEAL-safe parameters increase ciphertext size;  
(iii) \emph{masking leakage} — the overflow fix yields small distributional differences that could leak match information across repeated queries.
Overall, the results show that the 5G-aligned protocol remains practical with conservative BFV parameters and bounded masking, supporting homomorphic identifier verification within 5G latency requirements.

\section*{Acknowledgement}
The authors were financed based on the budget passed by the Saxonian State Parliament in Germany.
Additionally, this work has been partly funded by the German Federal Office for Information Security (BSI) project 6G-ReS (grant no. 01MO23013D). 
\bibliographystyle{IEEEtran}
\bibliography{references}

@inproceedings{mahdavi_pepsi_2024,
	location = {{USA}},
	title = {{PEPSI}: practically efficient private set intersection in the unbalanced setting},
	isbn = {978-1-939133-44-1},
	series = {{SEC} '24},
	shorttitle = {{PEPSI}},
	pages = {6453--6470},
	booktitle = {Proceedings of the 33rd {USENIX} Conference on Security Symposium},
	publisher = {{USENIX} Association},
	author = {Mahdavi, Rasoul Akhavan and Lukas, Nils and Ebrahimianghazani, Faezeh and Humphries, Thomas and Kacsmar, Bailey and Premkumar, John and Li, Xinda and Oya, Simon and Amjadian, Ehsan and Kerschbaum, Florian},
	urldate = {2025-02-22},
	date = {2024-08-12}
}

@article{el-hajj_evaluating_2024,
	title = {Evaluating the Efficiency of zk-{SNARK}, zk-{STARK}, and Bulletproof in Real-World Scenarios: A Benchmark Study},
	volume = {15},
	rights = {http://creativecommons.org/licenses/by/3.0/},
	issn = {2078-2489},
	url = {https://www.mdpi.com/2078-2489/15/8/463},
	doi = {10.3390/info15080463},
	shorttitle = {Evaluating the Efficiency of zk-{SNARK}, zk-{STARK}, and Bulletproof in Real-World Scenarios},
	pages = {463},
	number = {8},
	journaltitle = {Information},
	author = {El-Hajj, Mohammed and Oude Roelink, Bjorn},
	urldate = {2025-02-05},
	date = {2024-08},
	langid = {english},
	note = {Number: 8
Publisher: Multidisciplinary Digital Publishing Institute}
}

@incollection{cox_architecture_2021,
	title = {Architecture of the Core Network},
	isbn = {978-1-119-60268-2},
	url = {https://onlinelibrary.wiley.com/doi/abs/10.1002/9781119602682.ch2},
	pages = {29--53},
	booktitle = {An Introduction to 5G},
	publisher = {John Wiley \& Sons, Ltd},
	author = {Cox, Christopher},
	urldate = {2025-01-27},
	date = {2021},
	langid = {english},
	doi = {10.1002/9781119602682.ch2},
	note = {Section: 2
\_eprint: https://onlinelibrary.wiley.com/doi/pdf/10.1002/9781119602682.ch2}
}

@misc{3gpp_system_2024,
  author    = {3GPP},
  title     = {{3GPP TS 23.501 V19.1.0: System architecture for the 5G System (5GS)}},
  version   = {V19.1.0},
  number    = {{TS} 23.501},
  publisher = {3GPP},
  url       = {https://www.3gpp.org/ftp/Specs/archive/23_series/23.501/},
  date      = {2024-09}
}

@misc{3gpp_5g_2024-4,
  author    = {3GPP},
  title     = {{3GPP TS 29.571 V19.0.0: 5G System; Common Data Types for Service Based Interfaces; Stage 3}},
  version   = {V19.0.0},
  number    = {{TS} 29.571},
  publisher = {3GPP},
  url       = {https://www.3gpp.org/ftp/Specs/archive/29_series/29.571/},
  date      = {2024-09}
}

@misc{3gpp_numbering_2024,
  author    = {3GPP},
  title     = {{3GPP TS 23.003 V18.6.0: Numbering, addressing and identification}},
  version   = {V18.6.0},
  number    = {{TS} 23.003},
  publisher = {3GPP},
  url       = {https://www.3gpp.org/ftp/Specs/archive/23_series/23.003/},
  date      = {2024-06}
}

@misc{3gpp_procedures_2024,
  author    = {3GPP},
  title     = {{3GPP TS 23.502 V19.1.0: Procedures for the 5G System (5GS)}},
  version   = {V19.1.0},
  number    = {{TS} 23.502},
  publisher = {3GPP},
  url       = {https://www.3gpp.org/ftp/Specs/archive/23_series/23.502/},
  date      = {2024-09}
}

@misc{3gpp_5g_2024-1,
  author    = {3GPP},
  title     = {{3GPP TS 29.511 V18.3.0: 5G System; Equipment Identity Register Services; Stage 3}},
  version   = {V18.3.0},
  number    = {{TS} 29.511},
  publisher = {3GPP},
  url       = {https://www.3gpp.org/ftp/Specs/archive/29_series/29.511/},
  date      = {2024-06}
}

@misc{gsma_ts06_2024,
	title = {{TS}.06 - {IMEI} Allocation and Approval Process},
	url = {https://www.gsma.com/get-involved/working-groups/wp-content/uploads/2024/07/TS.06-v26.0-IMEI-Allocation-and-Approval-Process.docx},
	shorttitle = {{TS}.06},
	publisher = {{GSM} Association},
	author = {{GSMA}},
	urldate = {2024-12-28},
	date = {2024-07-01}
}

@inproceedings{wang2021privacy,
  title={$\{$Privacy-Preserving$\}$ and $\{$Standard-Compatible$\}$$\{$AKA$\}$ Protocol for 5G},
  author={Wang, Yuchen and Zhang, Zhenfeng and Xie, Yongquan},
  booktitle={30th USENIX security symposium (USENIX security 21)},
  pages={3595--3612},
  year={2021}
}

@inproceedings{koutsos20195g,
  title={The 5G-AKA authentication protocol privacy},
  author={Koutsos, Adrien},
  booktitle={2019 IEEE European symposium on security and privacy (EuroS\&P)},
  pages={464--479},
  year={2019},
  organization={IEEE}
}

@article{you20235g,
  title={5G-AKA-FS: A 5G authentication and key agreement protocol for forward secrecy},
  author={You, Ilsun and Kim, Gunwoo and Shin, Seonghan and Kwon, Hoseok and Kim, Jongkil and Baek, Joonsang},
  journal={Sensors},
  volume={24},
  number={1},
  pages={159},
  year={2023},
  publisher={MDPI}
}

@inproceedings{eleftherakis2024demystifying,
  title={Demystifying privacy in 5g stand alone networks},
  author={Eleftherakis, Stavros and Otim, Timothy and Santaromita, Giuseppe and Zayas, Almudena Diaz and Giustiniano, Domenico and Kourtellis, Nicolas},
  booktitle={Proceedings of the 30th Annual International Conference on Mobile Computing and Networking},
  pages={1330--1345},
  year={2024}
}

@article{scalise2025survey,
  title={A Survey of 5G Core Network User Identity Protections, Concerns, and Proposed Enhancements for Future 6G Technologies},
  author={Scalise, Paul and Hempel, Michael and Sharif, Hamid},
  journal={Future Internet},
  volume={17},
  number={4},
  pages={142},
  year={2025},
  publisher={MDPI}
}

@inproceedings{schmitt2021pretty,
  title={Pretty good phone privacy},
  author={Schmitt, Paul and Raghavan, Barath},
  booktitle={30th USENIX Security Symposium (USENIX Security 21)},
  pages={1737--1754},
  year={2021}
}

@inproceedings{intoci2023p3li5,
  title={P3LI5: practical and confidential lawful interception on the 5G core},
  author={Intoci, Francesco and Sturm, Julian and Fraunholz, Daniel and Pyrgelis, Apostolos and Barschel, Colin},
  booktitle={2023 IEEE Conference on Communications and Network Security (CNS)},
  pages={1--9},
  year={2023},
  organization={IEEE}
}

@manual{DoT_CEIR_2023v02,
  title        = {Central Equipment Identity Register (CEIR) Mobile User Manual – English v02},
  author       = {{Department of Telecommunications, Government of India}},
  organization = {Ministry of Communications},
  year         = {2023},
  version      = {v02},
  url          = {https://www.ceir.gov.in/HelpDocuments/English/CEIR_Mobile_User_Manual_English_v02.pdf},
  note         = {Accessed 2025-11-02}
}

@techreport{3GPP_TS29511_2023,
  title        = {TS 29.511: 5G System; Equipment Identity Register Services; Stage 3 (Release 17)},
  author       = {{3rd Generation Partnership Project (3GPP)}},
  institution  = {3GPP},
  year         = {2023},
  month        = jun,
  url          = {https://www.3gpp.org/ftp/Specs/archive/29_series/29.511/},
  note         = {Version: Release 17, accessed 2025-11-02}
}

@article{saeed2022comprehensive,
  title={A comprehensive review on the users’ identity privacy for 5G networks},
  author={Saeed, Mamoon M and Hasan, Mohammad Kamrul and Obaid, Ahmed J and Saeed, Rashid A and Mokhtar, Rania A and Ali, Elmustafa Sayed and Akhtaruzzaman, Md and Amanlou, Sanaz and Hossain, AKM Zakir},
  journal={IET Communications},
  volume={16},
  number={5},
  pages={384--399},
  year={2022},
  publisher={Wiley Online Library}
}

\end{document}